\documentclass[aps,12pt,showpacs]{revtex4}

\usepackage{epsfig}

\newcounter{fig}

\begin{document}
\title{\Large The cosmology of the nonsymmetric theory of gravitation}

\author{\Large Tomislav Prokopec$^*$ and Wessel Valkenburg}

\email[]{T.Prokopec@phys.uu.nl,w.valkenburg@phys.uu.nl}

\bigskip

\affiliation{(2) Institute for Theoretical Physics (ITP) \& Spinoza Institute,
             Minnaertgebouw, Leuvenlaan 4, Utrecht University
             3584 CE Utrecht, The Netherlands}
\vskip 0.1cm

\date{\today}

\begin{abstract}
 We show that during cosmological inflation the nonsymmetric metric
tensor theory of gravitation develops a spectrum which is potentially
observable by cosmic microwave background observations, and may be 
the most sensitive probe of the scale of cosmic inflation.
\end{abstract}


\maketitle


{\it 1. Introduction.}
 The metric tensor of the general relativistic theory of gravitation
is well known to be symmetric under the exchange of indices,
$g_{\mu\nu}(x)=g_{\nu\mu}(x) \equiv g_{(\mu\nu)}(x)$, where
$ g_{(\mu\nu)} = (1/2)(g_{\mu\nu} + g_{\nu\mu})$ denotes the 
symmetric part of the metric tensor. No physical principle precludes us 
from considering a more general theory of gravitation, in which the metric 
tensor, $g_{\mu\nu}\rightarrow \bar g_{\mu\nu}$,
contains a small antisymmetric admixture~\cite{Moffat:1979}, 
\begin{equation}
  \bar g_{\mu\nu}(x) =  g_{\mu\nu}(x) + B_{\mu\nu}(x)
\,,\qquad g_{\mu\nu} = \bar g_{(\mu\nu)}
\,,
\end{equation}
where $B_{\mu\nu} =
  \bar g_{[\mu\nu]} = (1/2) (\bar g_{\mu\nu}-\bar g_{\nu\mu})$ 
denotes the antisymmetric part of the metric tensor. 
This type of generalisation of 
the general relativistic theory of gravitation was first considered by 
Einstein~\cite{Einstein:1925}, 
in an attempt to unify gravitation with electromagnetism, whereby
$B_{\mu\nu}$ was interpreted as the electromagnetic field strength tensor. 
 From the tests of the Einstein theory of gravitation~\cite{Will:2001},
we know that the antisymmetric part of the metric tensor is small, and hence
the action can be well approximated by the Einstein-Hilbert action, plus 
a linearised antisymmetric 
contribution~\cite{DamourDeserMcCarthy:1992},
\begin{eqnarray}
 {\cal S} &=&  {\cal S}_{\rm EH} + S_{\rm NGT}
\label{action}
\\
{\cal S}_{\rm EH} &=& - \frac{1}{16\pi G_N}\int d^4x \sqrt{-g}\,
                       ({\cal R} +2\Lambda)
\label{action:EH}
\\
{\cal S}_{NGT} &=& \int d^4x \sqrt{-g}\, \Big(\frac{1}{12}
                     g^{\mu\alpha}g^{\nu\beta}g^{\rho\gamma}
                   H_{\mu\nu\rho}H_{\alpha\beta\gamma}
\nonumber\\
       &&\hskip 1.8cm   -\, \frac 14 \, m_B^2 g^{\mu\alpha}g^{\nu\beta}
                         B_{\mu\nu}B_{\alpha\beta}\Big)
\,,\quad
\label{action:NGT}
\end{eqnarray}
where ${\cal R}$ denotes the Ricci scalar,
$\Lambda$ the cosmological term, $G_N$ the Newton constant, and
\begin{equation}
 H_{\mu\nu\rho} = \partial_\mu B_{\nu\rho}
                + \partial_\nu B_{\rho\mu} 
                + \partial_\rho B_{\mu\nu}
\,
\label{field strength}
\end{equation}
is the field strength associated with the antisymmetric tensor 
(Kalb-Ramond) field $B_{\mu\nu}$. 
We work in natural units, in which $c=1$ and $\hbar = 1$. 

 In Eq.~(\ref{action:NGT}) we added a mass term for stability
reasons~\cite{DamourDeserMcCarthy:1991+1993,DamourDeserMcCarthy:1992}.
Moffat has recently argued~\cite{Moffat:2004} 
(see also~\cite{Moffat:1994}) that the Einstein theory 
plus a small massive antisymmetric component provides a viable explanation 
for the missing matter problem of the Universe, which is standardly cured 
by adding a dark (nonbaryonic) matter of unknown composition and origin.
The inverse mass scale, $m_B^{-1}$ gives the scale at which the effective 
strength of the gravitational interaction changes. At distances smaller than 
$m_B^{-1}$ the Newton force is equal to the observed value, at distances 
larger than $m_B^{-1}$ the Newton force is stronger, 
explaining thus the rotation curves of galaxies, as well
as the gravitational lensing of light by galaxies and clusters of galaxies.
Note that when the mass term in~(\ref{action:NGT}) is nonzero, 
the NGT theory~(\ref{action:NGT}) ceases to be equivalent to 
the Kalb-Ramond axion~\cite{GasperiniVeneziano:2002}. 

 In this Letter we consider the cosmological aspects of the nonsymmetric
theory of gravitation. By canonically quantising
the physical components of the nonsymmetric tensor field during inflation,
we study the growth of quantum fluctuations during
inflation~\cite{MukhanovChibisov:1981},
 and then evolve them during radiation and matter eras.
The nonsymmetric tensor field appears naturally in flux compactifications
in string theory, where it is disguised as the Kalb-Ramond axion, whose 
cosmological relevance has been studied in 
detail~\cite{GasperiniVeneziano:2002,VernizziMelchiorriDurrer:2000}.


\vskip 0.1in


{\it 2. Conformal space-times.}
 The metric tensor of spatially homogeneous conformal space times,
which include cosmic inflation and Friedmann-Lema\"itre-Robertson-Walker
(FLRW) space-times, has the form
\begin{equation}
 g_{\mu\nu} = a^2(\eta)\eta_{\mu\nu}
\,,
\label{metric tensor}
\end{equation}
where $\eta_{\mu\nu} = {\rm diag}(1,-1,-1,-1)$ denotes the Minkowski metric, 
and $a=a(\eta)$ is the conformal (scale) factor. 

 In conformal space-times the action of the nonsymmetric tensor 
theory~(\ref{action:NGT}) simplifies to,
\begin{eqnarray} 
 {\cal S}_{NGT} \rightarrow 
    {\cal S}_{NGT}^{\rm conf} &=& \int d^4x\, \Big(\frac{1}{12}\frac{1}{a^2}
                   \eta^{\mu\alpha}\eta^{\nu\beta}\eta^{\rho\gamma}
                   H_{\mu\nu\rho}H_{\alpha\beta\gamma}
\nonumber\\
  &&\hskip 0.9cm    -\, \frac 14 \, m_B^2 \eta^{\mu\alpha}\eta^{\nu\beta}
                         B_{\mu\nu}B_{\alpha\beta}\Big)
\label{action:NGT:conformal}
.
\end{eqnarray}
Unlike vector gauge fields~\cite{ProkopecPuchweinWoodard:2003},
the antisymmetric tensor field 
does not couple conformally to gravitation.
The corresponding equation of motion is easily obtained by varying
the action~(\ref{action:NGT:conformal}),
\vskip -0.5cm
\begin{equation}
 \Big(\partial^2 + a^2 m_B^2 \Big) B_{\mu\nu}
  - 2\frac{a^\prime}{a}H_{0\mu\nu} = 0
\,,
\label{eom}
\end{equation}
where $\partial^2 \equiv \eta^{\mu\nu}\partial_\mu\partial_\nu$.
Note that the antisymmetric tensor field $B_{\mu\nu}$ 
is anti-damped by the Universe's expansion.
Upon taking the divergence $\eta^{\mu\alpha}\partial_\alpha$ 
of~(\ref{eom}) divided by $a^2$, we arrive at the following 
consistency (gauge) condition,
\vskip -0.8cm
\begin{equation}
 \eta^{\mu\nu}\partial_\mu B_{\nu\rho} = 0
\,,
\label{Lorentz condition}
\end{equation}
which is analogous to the Lorentz gauge condition of electromagnetism.
Equations~(\ref{eom}) and~(\ref{Lorentz condition}) fully specify the dynamics 
of a massive antisymmetric tensor field in conformal space-times. 

 We now make use of the following electric/magnetic
decomposition of the antisymmetric tensor field,
\vskip -0.5cm
\begin{equation}
 B_{0i} \!=\! E_i \!=\! - B_{i0}
\,,\;\,  B_{ij} \!=\! -\epsilon_{ijl} B_{l}
\; (i,j,l=1,2,3)
\,,
\label{electric magnetic decomposition}
\end{equation}
where $\epsilon_{ijl}$  
denotes the totally antisymmetric symbol 
($\epsilon_{123} = 1, \epsilon_{321} = -1$, {\it etc.}).
With this decomposition, Eq.~(\ref{eom}) becomes
\vskip -0.6cm
\begin{eqnarray}
 \Big(\partial^2 + a^2 m_B^2 \Big) \vec E &=& 0
\label{eom:E}
\\
 \Big(\partial^2 + a^2 m_B^2 \Big) \vec B
  - 2\frac{a^\prime}{a}\Big(\partial_\eta \vec B
                          + \vec \partial \times \vec E
                      \Big) &=& 0
\,,\qquad
\label{eom:B}
\end{eqnarray}
while the Lorentz condition~(\ref{Lorentz condition}) implies 
the following `constraint' equations
\vskip -0.6cm
\begin{eqnarray}
\quad
 \vec \partial\cdot \vec E &=& 0
\label{eom:E constraint}
\\
 \vec \partial_\eta \vec E - \vec \partial \times \vec B &=& 0
\,.
\label{eom:B constraint}
\end{eqnarray}
Note that these four equations are equivalent to the three vacuum Maxwell 
equations~(\ref{eom:E constraint}), (\ref{eom:B constraint}), and 
$\partial_\eta \vec B + \vec \partial \times \vec E = 0$. 
An important difference with respect to the Maxwell theory is 
that the constraint equation,
 $\vec \partial \cdot \vec B = 0$, is missing. This then implies
that, unlike in the Maxwell theory, the longitudinal magnetic component
$\vec B^L$ of the antisymmetric tensor field is dynamical. 
Equations~(\ref{eom:B}--\ref{eom:B constraint}) imply that, in the massive 
case, the transverse electric component, $\vec E^T$,
and the longitudinal magnetic component, $\vec B^L$, 
comprise the three physical degrees of freedom,
while $\vec E^L = 0 $ and $\vec B^T$ is specified
in terms of $\vec E^T$, as given in~(\ref{eom:B constraint}).
In the massless theory however, there is a remaining 
gauge freedom~\cite{KalbRamond:1974}. 
Similarly to the gauge fixing, $A_0 = 0$, in electromagetism, 
one can choose the gauge, $\vec E^T=0$. This then implies that 
the longitudinal magnetic component, $\vec B^L$,
is the only remaining physical degree of freedom 
of the massless nonsymmetric tensor theory.

\vskip 0.1in


{\it 3. De Sitter inflation.}
 Let us now consider de Sitter inflation, in which the scale factor  
is a simple function of conformal time $\eta$, 
\vskip -0.5cm
\begin{equation}
 a=-\frac{1}{H_I\eta}
\,\qquad (\eta\leq -1/H_I)
\,\qquad {\tt (de\; Sitter)}
\,,
\label{scale factor:de Sitter}
\end{equation}
and $H_I$ is the Hubble parameter during inflation. 
Let us for the moment assume that the mass scale $m_B^{-1}$ is larger than 
any relevant physical scale in the theory, such that it can be omitted  
from Eqs.~(\ref{eom:E}--\ref{eom:B}).
 In this limit the transverse electric component couples conformally
to gravitation, such that its evolution corresponds to that of conformal
vacuum, whith the identical correlations
as the Minkowski vacuum of gauge fields, 
and hence are of no relevance for cosmology. 

 We therefore focus on the longitudinal magnetic component.
We now perform a canonical quantisation,
\begin{eqnarray}
 \hat{\vec B^L} = a\!\!\int\! \frac{d^3 k}{(2\pi)^3}
                           {\rm e}^{i\vec k\cdot \vec x}
                           \vec \epsilon^{\,L}\!(\vec k)
                          \Big[
                               B^L_{\vec k}(\eta) \hat b_{\vec k}   
                   + {B^L_{-\vec k}}\!\!^*(\eta) \hat b^\dagger_{\!-\vec k}
                           \,\Big]
,\;\;
\label{canonical quantisation:BL}
\end{eqnarray}
with $\big[\hat b_{\vec k}, \hat b^\dagger_{\vec k^\prime} \big]
                  =   (2\pi)^3\delta(\vec k-\vec k^{\;\prime}\,)$.
The (conformally rescaled) 
mode functions obey Eq.~(\ref{eom:B}), with $m_B=0$,
\begin{eqnarray}
 \Big(
      \partial_\eta^2+\vec k^2 + \frac{a^{\prime\prime}}{a}
                               - 2\Big(\frac{a^\prime}{a}\Big)^2
 \Big)  B^L_{\vec k}(\eta) &=& 0
\,.
\label{eom:BL:2}
\end{eqnarray}
In de Sitter inflation~(\ref{scale factor:de Sitter}), 
${\hat{\vec B^L}}(\vec x,\eta)/a$, couples conformally, 
\begin{equation}
 \big(
      \partial_\eta^2+\vec k^2
 \big)  B^L_{\vec k}(\eta) = 0
\,\qquad {\tt (de\; Sitter\; era)}
\,,
\label{eom:BL:3}
\end{equation}
implying the following amplitude of vacuum fluctuations,
$
  B^L_{\vec k}(\eta) =  (2k)^{-1/2}\, {\rm e}^{-i k\eta}
$,
where $k=\|\vec k\|$, $\vec \epsilon^{\,L}(\vec k)$ is the longitudinal 
polarization vector, $\vec k\times \vec \epsilon^{\,L}(\vec k) =0 $, 
and the Wronskian reads, 
${\bf W}\big[B^L_{\vec k}(\eta),{B^L}^*_{\vec k}(\eta)\big] = i$.
Hence, during de Sitter inflation, 
the physical field, $\vec B^{\,L}$,
exhibits conformal vacuum correlations.

\vskip 0.1in


{\it 4. Radiation and Matter Era.}
Let us now consider radiation and matter era, in which
the scale factors read,
\vskip -0.5cm
\begin{eqnarray}
  a &=& H_I \eta
\,\quad (\eta_{\rm e}\geq \eta \geq 1/H_I)
\,\quad {\tt (radiation\;era)}
\, 
\label{scale factor:radiation}
\quad\,\,
\\
  a &=& \frac{H_I}{4\eta_{\rm e}}\Big(\eta + \eta_{\rm e}\Big)^2
\,\quad (\eta \geq \eta_{\rm e})
\,\quad {\tt (matter\;era)}
\, ,
\label{scale factor:matter}
\end{eqnarray}
where we assumed a sudden radiation-to-matter transition.
$\eta_{\rm e}=(H_IH_{\rm e})^{-1/2}$ denotes the conformal time at
the matter and radiation equality, 
and it is defined by, $a_0/a_{\rm e} \equiv 1+z_{\rm e} 
           = (\eta_0/\eta_{\rm e})^2$,
where $a_0=a(\eta_0)$ denotes the scale factor today, and 
$z_{\rm e} = 3230\pm 200$~\cite{Spergeletal:2003WMAP} 
is the redshift at the radiation-matter equality.


In radiation era Eq.~(\ref{eom:BL:2}) reduces to 
\vskip -0.5cm
\begin{equation}
  \Big(
       \partial_\eta^2+\vec k^2 - \frac{2}{\eta^2}
  \Big) B^L_{\vec k}(\eta) = 0
\,.
\label{eom:BL:radiation era}
\end{equation}
This is the Bessel equation with the index, $\nu = 3/2$, 
and whose general solution is a linear combination of Hankel functions,
$H_{3/2}^{(1)}(k\eta)$ and $H_{3/2}^{(2)}(k\eta)$ 
(analogous to the Bunch-Davies vacuum in de Sitter space),
\begin{equation}
 B^L_{\vec k}(\eta) = \frac{1}{\sqrt{2k}}
                   \Big[
                        \alpha_{\vec k}\Big(1\!-\!\frac{i}{k\eta}\Big)
                            {\rm e}^{-ik\eta}
                      +  \beta_{\vec k}\Big(1\!+\!\frac{i}{k\eta}\Big)
                            {\rm e}^{ik\eta}
                   \Big]
\,.
\label{BL:radiation era}
\end{equation}
Upon choosing the coefficients $\alpha_{\vec k}$ and $\beta_{\vec k}$ such
that $|\alpha_{\vec k}|^2-|\beta_{\vec k}|^2 = 1$, the Wronskian 
becomes canonical, 
${\bf W}\big[B^L_{\vec k}(\eta),{B^L}^*_{\vec k}(\eta)\big] = i$.
Note that
the mode amplitude~(\ref{BL:radiation era})
exhibits a $1/k$ infrared enhancement in radiation era
on superhubble scales.

 This enhancement leads to mode mixing at the inflation-radiation transition.
Indeed, upon performing a continuous matching of $B_{\vec k}^{\,L}$ and
$\partial_\eta B_{\vec k}^{\,L}$ at the inflation-radiation 
transition, we arrive at 
\begin{equation} 
 \alpha_{\!\vec k} \!=\! - \frac{1}{2}\frac{H_I^2}{k^2}
                  \Big[
                       1 - 2i \frac{k}{H_I} - 2\Big(\frac{k}{H_I}\Big)^2\,
                  \Big]{\rm e}^{2ik/H_I}
\!\!,\;
 \beta_{\vec k} = - \frac{1}{2}\frac{H_I^2}{k^2}
,
\label{alpha-beta}
\end{equation} 
such that for superhubble modes
at the end of inflation,
%
\begin{equation} 
\beta_{\vec k}  \simeq \alpha_{\vec k} = -\frac{H_I^2}{2k^2}
\, \qquad  (k\ll H_I)
\,.
\label{alpha-beta:superhubble}
\end{equation} 

\vskip 0.2cm

On the other hand, in matter era~(\ref{scale factor:matter}),
Eq.~(\ref{eom:BL:2}) becomes
\vskip -0.6cm
\begin{equation}
  \Big(
       \partial_\eta^2+\vec k^2 - \frac{6}{(\eta+\eta_{\rm e})^2}
  \Big) B^L_{\vec k}(\eta) = 0
\,.
\label{eom:BL:matter era}
\end{equation}
The fundamental solutions are proportional to 
$H_{5/2}^{(1)}\big(k(\eta+\eta_{\rm e})\big)$ 
and $H_{5/2}^{(2)}(k(\eta+\eta_{\rm e})\big)$. More precisely,
\begin{eqnarray}
 B^L_{\vec k}(\eta) \!\!&=&\!\! \frac{1}{\sqrt{2k}}
                   \Big[
                        \gamma_{\vec k}
                        \Big(1-\frac{3i}{k\tilde\eta}-\frac{3}{(k\tilde\eta)^2}
                        \Big)
                            {\rm e}^{-ik\tilde\eta}
\label{BL:matter era}
\\
        &&\hskip 0.6cm +\,  \delta_{\vec k}
                        \Big(1+\frac{3i}{k\tilde\eta}-\frac{3}{(k\tilde\eta)^2}
                        \Big)
                            {\rm e}^{ik\tilde\eta}
                   \Big]
\,\quad {\tt (matter)}
\nonumber\,,
\end{eqnarray}
with $|\gamma_{\vec k}|^2-|\delta_{\vec k}|^2 = 1$ and 
$\tilde\eta=\eta+\eta_{\rm e}$. Note that 
the mode amplitude~(\ref{BL:matter era})
exhibits a $1/k^2$ infrared enhancement on superhubble scales. 

 We now continuously match $B_{\vec k}^{\,L}$ and
$\partial_\eta B_{\vec k}^{\,L}$ at the radiation-matter 
transition, $\eta=\eta_{\rm e}$ ($\tilde\eta=2\eta_{\rm e}$), to get  
\begin{eqnarray} 
 \gamma_{\vec k}{\rm e}^{\!\!-ik\tilde\eta_{\rm e}} \!\!\!&=&\!\! 
                   \alpha_{\vec k}
                   \Big(\!1
                      \!+\!  \frac12\frac{i}{k\eta_{\rm e}}
                      \!-\! \frac18\frac{1}{(k\eta_{\rm e})^2}
                   \Big){\rm e}^{\!-ik\eta_{\rm e}}
                 + \beta_{\vec k}
                   \frac18\frac{{\rm e}^{ik\eta_{\rm e}}}{(k\eta_{\rm e})^2}
\,,\qquad 
\nonumber\\
 \delta_{\vec k}{\rm e}^{ik\tilde\eta_{\rm e}} \!\!\!&=&\!\!
                   \alpha_{\vec k}
                   \frac18\frac{{\rm e}^{\!-ik\eta_{\rm e}}}{(k\eta_{\rm e})^2}
                    + \beta_{\vec k}
                   \Big(\!
                        1
                      \!-\! \frac12\frac{i}{k\eta_{\rm e}}
                      \!-\! \frac18\frac{1}{(k\eta_{\rm e})^2}
                   \Big){\rm e}^{ik\eta_{\rm e}}
\,.
\nonumber
\end{eqnarray} 
%


\vskip 0.1in


{\it 5. The spectrum.}
%
Since we are interested in how 
a nonsymmetric tensor field $B_{\mu\nu}$ may affect 
large scale structures of the Universe, 
the spectrum of $B_{\mu\nu}$ can be defined
(in analogy to matter density and magnetic field 
perturbations~\cite{ProkopecPuchwein:2004})
in terms of the corresponding stress-energy tensor,
 ${T_{\mu\nu}}_{\rm NGT} = [2/\sqrt{-g}]
                            \delta S_{NGT}/\delta g^{\mu\nu}$.
When expressed in terms of the electric and magnetic components, 
and setting $m_B\rightarrow 0$, one finds,
\begin{equation}
  {T_{0}^{\;0}}_{\rm NGT} = \frac{1}{2a^6} 
                     \Big[
                          (\partial_\eta \vec B + \vec \partial\times\vec E)^2
                        + (\vec \partial\cdot \vec B)^2
                     \Big] \equiv \rho_{\rm NGT}
\,,
\label{T00}
\end{equation}
where $\rho_{\rm NGT}$ denotes the energy density of $\vec B^L$. 
Transforming into momentum space, 
we then get for the spectrum~({\it cf.} Ref.~\cite{ProkopecPuchwein:2004}),
\begin{eqnarray}
 {\cal P}_{\rm NGT} &=& \frac{k^3}{2\pi^2}{T_{0}^{\;0}}^{\rm NGT}(\vec k,\eta) 
\nonumber
\\ 
  &=& \frac{k^3}{4\pi^2 a^4}
   \Big[
         \Big|\partial_\eta B_{\vec k}^L+\frac{a^\prime}{a}B_{\vec k}^L\Big|^2 
      +  k^{\,2}
|B_{\vec k}^L|^2
   \Big]                    
\,.\quad
\label{spectrum}
\end{eqnarray}
%

%
\begin{figure}[tbp]
\vskip -0.2in
\centerline{\hspace{-0.1in} 
\epsfig{file=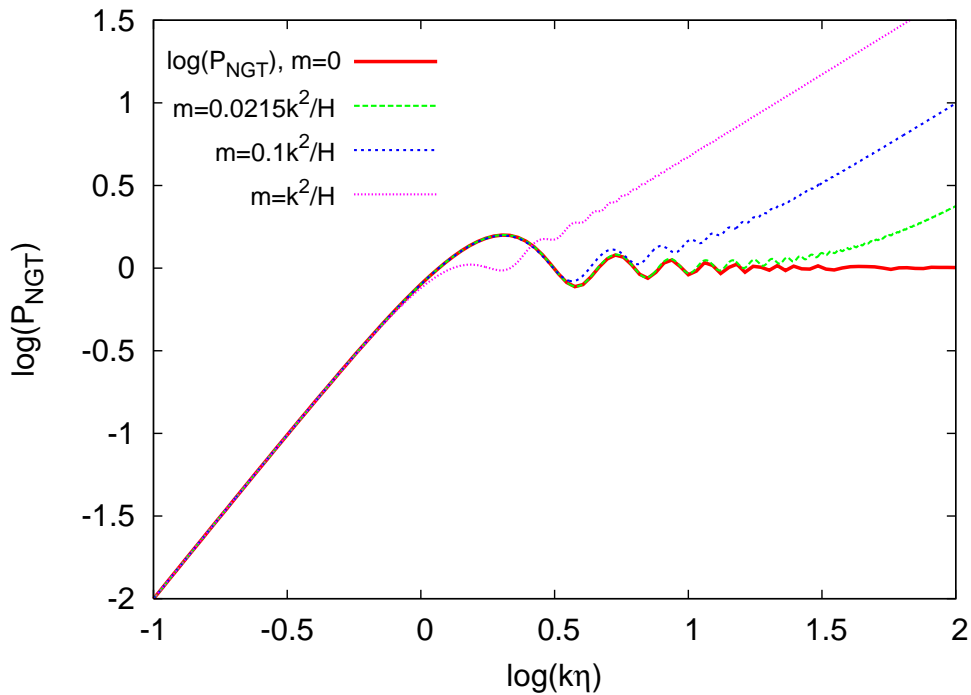, width=5.3in}
}
\vskip -0.1in
\caption{\small
The spectrum of the longitudinal magnetic component of a massive
nonsymmetric tensor theory in radiation era. 
The spectrum is of the form,
 ${\cal P}_{\rm NGT}\propto k^2$, on superhubble scales, 
and it reduces to a constant plus a decaying oscillating component on 
subhubble scales ($m=0$, solid red). 
On scales $k/a\ll m$ the spectrum exhibits a linear growth with time
(green, blue, violet).
}
\vskip -0.1in
\label{figure one}
\end{figure}
 To obtain the spectrum in radiation era, we insert the mode 
functions~(\ref{BL:radiation era}) into~(\ref{spectrum}), with the matching 
coefficients given in~(\ref{alpha-beta:superhubble}). The result is,
\begin{equation}
 {\cal P}_{\rm NGT}^{\tt rad} = \frac{H_I^4}{8\pi^2 a^4}
                      \bigg\{
                                 1+\frac12\frac{1}{(k\eta)^2}
                        -\frac12\frac{\cos(2k\eta)}{(k\eta)^2}
                        - \frac{\sin(2k\eta)}{k\eta}
                      \bigg\}
\label{spectrum:radiation era}
\,.
\end{equation}
In figure~\ref{figure one} we plot $\log[{\cal P}_{\rm NGT}]$ 
as a function of $\log(k\eta)$. Note that in the infrared $(k\eta\ll 1)$ 
the spectrum scales as, 
${\cal P}_{\rm NGT}\propto k^2$ (dashed red line),
which is enhanced with respect
to the inflationary spectrum,  ${\cal P}_{\rm NGT}\propto k^4$.
On subhubble scales $(k\eta\gg 1)$ the massless field spectrum reduces 
to a constant (solid red line), 
${\cal P}_{\rm NGT} \simeq H_I^4/(8\pi^2a^4)$ ($m=0$).
This amplitude is the same as that of gravitational wave fluctuations, 
and scales as a relativistic matter, plus a decaying oscillating component.
The spectrum of the massive field (longitudinal magnetic component)
exhibits in addition linear growth
on large scales, $k/a \ll m_B$, and the spectrum for various masses 
is shown in figure~\ref{figure one} (as a function of increasing $m_BH/k^2$).
 This implies that at any given time there is an
enhancement in power on large scales ($k/a \ll m_B$) 
by a factor $m_Ba/k$ when compared with the power in gravitational 
waves, rendering a massive nonsymmetric field potentially a more sensitive
probe of inflationary scale than gravitational waves. 
(The transverse components in~(\ref{eom:E})
do not exhibit a significant amplification, 
and hence we do not discuss them here.)
Since in a geometric theory, $m_B^2 \sim \Lambda$, an observation of 
any imprint in cosmic microwave background
may be a signal for a genuine cosmological term. 

\begin{figure}[tbp]
\vskip -0.2in
\centerline{\hspace{-0.1in} 
\epsfig{file=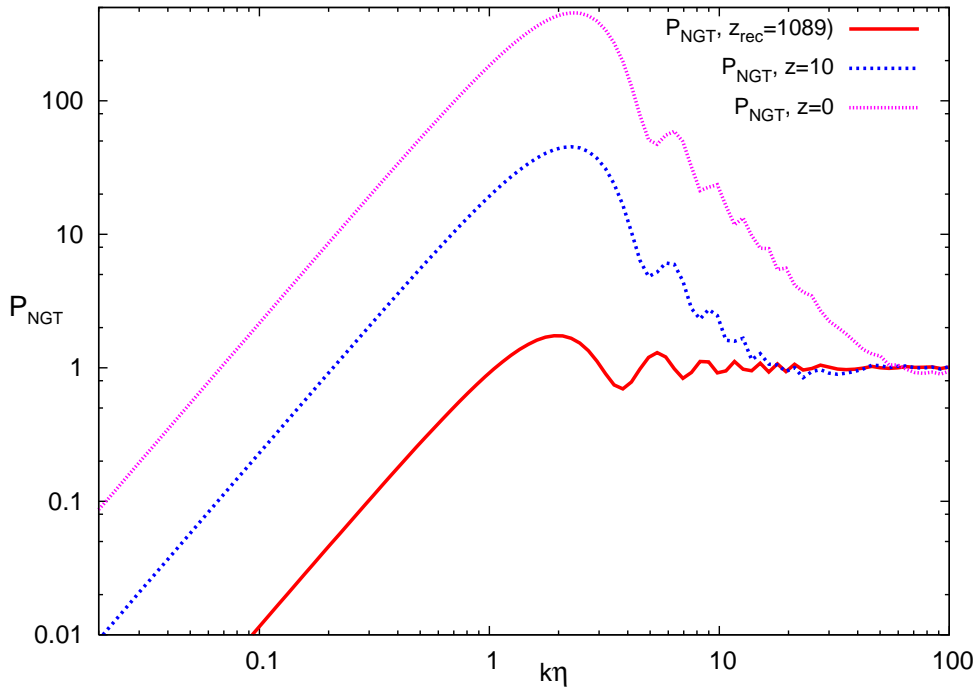, width=5.3in}
}
\vskip -0.1in
\caption{\small
The spectrum of the longitudinal magnetic component of the  massless
nonsymmetric tensor theory in matter era (log-log plot). 
We show the spectrum at recombination,
 $\eta=\eta_{\rm rec}$ ($z_{\rm rec} \simeq 1089$, solid red), 
structure formation, $\eta=\eta_{\rm 10}$ ($z\simeq 10$, dashed blue), 
and today $\eta=\eta_{\rm 0}$ ($z= 0$, dot-dashed violet).
The hubble crossing is at $k\eta\simeq 1$. 
}
\vskip -0.15in
\label{figure two}
\end{figure}

A complete analytical expression for the spectrum~(\ref{spectrum}) 
in matter era, with the modes given
by~(\ref{BL:matter era}), is rather complicated.
To illustrate its main features, we plot the matter era spectrum
for a massless nonsymmetric field in figure~\ref{figure two}: 
(a) at the time of electron-proton recombination 
($z=z_{\rm rec} \simeq 1089$, solid red), (b) at the time
of structure formation ($z \simeq 10$, dashed blue) and (c) today 
($z=0$, dot-dashed violet). The main features of the spectrum are as 
follows. Up to a small oscillating correction on subhubble scales,
corresponding to the momenta,  $k\eta \gg [z_{\rm e}/z(\eta)]^{1/2}$
($z_{\rm e} = 3230\pm 200$), the spectrum in matter era stays flat, 
${\cal P} \simeq H_I^4/(8\pi^2a^4)$. On subhubble scales the nonsymmetric 
tensor field scales as nonrelativistic matter, implying an enhancement 
relative to the radiation era spectrum, which is
given by the ratio of the scale factors at the hubble crossing
 and at time $\eta$, $a(\eta)/a(k^{-1})$. This results in the 
characteristic feature seen in figure~\ref{figure two}, with
${\cal P}\propto 1/k^2$ ($1\ll k\eta \ll [z_{\rm e}/z(\eta)]^{1/2} $),
to which decaying oscillations are superimposed. 
%
%


\vskip 0.1in

{\it 6. Discussion.}
 How the antisymmetric tensor field
affects microwave background anisotropies, 
depends on the precise nature of the coupling to the photon field,
and can be realised either {\it via} a geodesic equation, 
or a direct coupling to an electromagnetic field~\cite{JanssenProkopec}. 

 In this Letter we consider cosmological implications of the 
nonsymmetric tensor theory, by canonically quantising the 
physical component of the nonsymmetric tensor field in de Sitter inflation,
and subsequently evolving it in radiation and matter era. 
We find that in the massless limit the relevant physical excitation
(the longitudinal magnetic component), 
exhibits an approximately scale invariant
(energy density) spectrum in radiation and matter era on subhubble scales, 
and an infrared-safe spectrum on superhorizon scales,
${\cal P}_{\rm NGT}\propto k^2$, in both radiation and matter eras.
The spectrum of a massive theory gets amplified on large scales 
($k/a \ll m_B$) such that the amplitude exceeds that of gravitational waves,
rendering the nonsymmetric theory of gravitation potentially 
the most sensitive probe of inflationary scale. 






\end{document}